\documentclass{article}

\usepackage[preprint]{neurips_2021}

\usepackage[utf8]{inputenc} 
\usepackage[T1]{fontenc}    
\usepackage{hyperref}       
\usepackage{url}            
\usepackage{booktabs}       
\usepackage{amsfonts}       
\usepackage{nicefrac}       
\usepackage{microtype}      
\usepackage{xcolor}         

\usepackage{graphicx}
\usepackage{caption}
\usepackage{subcaption}
\usepackage{svg}
\usepackage{booktabs} 
\usepackage{tabularx}
\graphicspath{{./images/}{.}}
\makeatletter
\def\input@path{{./images/}{./}}
\makeatother

\usepackage{svg}
\usepackage{times}
\usepackage{epsfig}
\usepackage{graphicx}
\usepackage{amsmath}
\usepackage{amssymb}
\usepackage{inputenc}
\usepackage{algorithm}
\usepackage{adjustbox}
\usepackage[page]{appendix}
\newtheorem{lemma}{Lemma}

\newcommand{\nlparagraph}[1]{\paragraph{#1}\mbox{}\newline}

\title{Realistic molecule optimization on a learned graph manifold}

\author{%
  Rémy Brossard \\
  \texttt{remy@anotherbrain.ai}
  \And
  Oriel Frigo \\
  \texttt{oriel@anotherbrain.ai}
  \And
  David Dehaene \\
  \texttt{david@anotherbrain.ai}
}

\begin{document}

\maketitle

\begin{abstract}
  Deep learning based molecular graph generation and optimization has recently been attracting attention due to its great potential for \emph{de novo drug design}. On the one hand, recent models are able to efficiently learn a given graph distribution, and many approaches have proven very effective to produce a molecule that maximizes a given score. On the other hand, it was shown by previous studies that generated optimized molecules are often unrealistic, even with the inclusion of mechanics to enforce similarity to a dataset of real drug molecules. In this work we use a hybrid approach, where the dataset distribution is learned using an autoregressive model while the score optimization is done using the Metropolis algorithm, biased toward the learned distribution. We show that the resulting method, that we call learned realism sampling (LRS), produces empirically more realistic molecules and outperforms all recent baselines in the task of molecule optimization with similarity constraints.
\end{abstract}
\section{Introduction}
After tremendous success in task automation, machine learning has recently been used to accelerate expert development. In the field of medical research, new tools to efficiently exploit knowledge database or to accelerate simulations by orders of magnitude can be of great aid. In that matter, the problem of drug design is a particularly challenging but important problem: Before a compound is commercialized, it is proposed as a candidate and will have to pass an extensive battery of tests and certifications on animals and humans. In current modern pipelines, only one candidate in ten thousand will make its way to the market, explaining the development cost (between 1 and 3 billion of dollars) and time (10 to 15 years) needed for a single new drug.

With a large quantity of freely available data, many recent works aiming at improving this ratio have been proposed by the community. Searching for a good molecule requires two things: a way to score its quality, and a method to search the ensemble of reasonable candidates. The former was addressed by supervised algorithms, trained to predict many molecular properties of interest, that could be efficiently used to accelerate the prediction while increasing its accuracy \citep{gilmer2017neural, wu2018moleculenet}. Notably, the use of graph neural networks seems to be particularly adapted to these tasks \citep{Hu2020, wu2018moleculenet}. Although the latter has also been explored extensively by different methods, some remaining recurring issues still limit their practical use. In particular, \citet{gao2020synthesizability} have shown that molecules generated by deep learning models are often unrealistic, and cannot be synthesized. 

Molecular data can be represented in many forms, but in this work we focus on molecules represented as graphs, with atoms seen as nodes and chemical bonds seen as edges. In literature, three main approaches can be found. 
Firstly, a simple representation learning method can be used to produce an embedding of molecules in a smooth space, in which standard optimization methods can be used, such as bayesian optimization \citep{jin2018junction}. This approach relies heavily on the quality of the embedding space, limiting the optimization performance. A second one is to explicitly train a model to produce molecules with given properties \citep{jin2018learning}. However, this requires a large quantity of data specific to the target properties, limiting the usefulness of such methods for the elaboration of drugs with \emph{new} effects. Finally, reinforcement learning (RL) methods have proven effective as a pure optimization method \citep{zhou2019optimization, you2018graph, shi2020graphaf}. Most works often add a component to enforce realism regarding a dataset, such as a discriminator or pretraining, as it is often easier for the model to learn the specifics of the score function, leading to the generation of unrealistic molecules. Although one can argue that proposing new ways to optimize a score can be interesting, these models are difficult to use in practice.

In this work, we propose a different approach to enforce molecule realism, combining machine learning and traditional importance sampling. More precisely, we train a graph autoregressive generative model to learn the distribution of reasonable molecules on the ZINC dataset \citep{irwin2005zinc}. We then alter a simple Metropolis sampling algorithm to propose molecules which are both optimized and realistic. The resulting method, that we refer to as Learned Realism Sampling (LRS) does not need score-specific data, produces realistic molecules when RL algorithms do not, and improves performance on all our baselines in the task of molecule \emph{improvement} on a variety of different scores.

\section{Related work}
\label{related}

\subsection*{Molecule generation}
This work focuses mainly on graph representations of molecules.
Molecular graphs do not have a well defined node ordering and their ensemble lacks a topology, making the problem of molecular graph generation difficult in most deep learning frameworks. Three main approaches can be found in literature. First, a standard generative framework such as a Variational Autoencoder \citep{Kingma2013VAE} can be used to generate the graph adjacency matrix \citep{simonovsky2018graphvae}. Although simple, these methods suffer from the fact that they are not node permutation invariant. A second popular approach relies on the junction tree representation \citep{jin2018junction, jin2018learning}. Briefly, it consists in heuristics to represent the molecule as a tree, ie to remove the cycles. A tree is considerably simpler than a molecular graph, and recurrent methods can be efficiently applied to it. Finally, autoregressive models can be used to generate a graph node-by-node and edge-by-edge \citep{popova2019molecularrnn, shi2020graphaf, you2018graph, zhou2019optimization, li2018multi-objective}. This has the advantage of breaking the symmetries introduced by the incomplete ordering of the nodes.

More broadly used than graph representations, SMILES and more recently SELFIES\citep{krenn2020self} are text-based representations of molecules, on which standard sequence models can be used. The main drawback of this approach is that neighboring atoms in the molecule can be far away in the representation. In this regard, we highlight the work of \citet{alperstein2019all} that solved the problem by using multiple SMILES representations of the molecule. Here, we also consider the work of \cite[SMILES LSTM]{segler2018generating} in section \ref{section: unconstrained optimisation} as another baseline.

\subsection*{Optimization}
A first approach for property optimization is to perform representation learning to embed molecules as vectors in a smooth space. Standard algorithms such as Bayesian optimization\citep[JTVAE]{jin2018junction}, gradient descent \citep[All SMILES]{alperstein2019all} or hill climbing \citep[SMILES LSTM]{segler2018generating} can then be used on those representations . Although efficient in principle, these algorithms are limited by the quality of the representation as well as by the ability of the model to reconstruct the original molecule from it.
A second approach is to train the model in a \emph{domain transfer} framework, that is to train the model to take a molecule with low score as input and give a similar molecule with high score as output \citep[VJTNN or VJTNN+GAN]{jin2018learning}. These models can be effective, although they do not actually optimize properties. Notice that both these approaches rely heavily on the availability of labelled data. 

In order to overcome this last issue, a standard approach is to train a model to directly maximize a score with reinforcement learning. These algorithms are very efficient for the optimization itself. Methods to enforce molecule realism regarding a dataset are added, such as pretraining \citep[MRNN]{popova2019molecularrnn}\citep[GraphAF]{shi2020graphaf} or a discriminator \citep[GCPN]{you2018graph}. However, despite these efforts, they usually generate unrealistic samples, as highlighted in this work.

Finally, there is a proposition of an hybrid approach between machine learning and other algorithms. Most similar to our work, we highlight two recent models based on Markov Chain Monte Carlo (MCMC). The MARS model \citep{xie2021mars} trains and generates molecules simultaneously, adding better candidates to its training set. The generation procedure shares a lot of similarity with our work. Although the model produces impressive result in optimization, it does not attempt to enforce realism and we do not use it as a baseline. The MIMOSA model \citep{fu2020mimosa} performs MCMC and directly models the transition probability with a pretrained neural network, as opposed to this work, where we model a dataset distribution. 




\section{Method}
\subsection{Graph generation with an auto-regressive model}
\label{section:armodel}

We consider a dataset representing a distribution $\mathcal{Q}(G)$ over molecular graphs G that we wish to approximate. We first represent a graph $G$ in a constructive fashion, as a sequence $s$ of vectors which can either represent the addition of an atom or of a chemical bond, as detailed in section \ref{section:sequence definition}. The resulting distribution $\mathcal{Q}(s)$ is then approximated by an autoregressive distribution $\mathcal{P}(s)$ parameterized with neural networks, as detailed in section \ref{section: nn}.

\subsubsection{Graphs as sequences}
\label{section:sequence definition}
We consider a molecule, seen as a complete graph $G(\mathcal{V}, \mathcal{E})$ where $\mathcal{V} = 1 .. N$ is the set of nodes and $\mathcal{E}\subset \mathcal{V}\times\mathcal{V}$ is the set of \emph{undirected} edges, $x_v \in \{1..D\}$ is the label of the node $v\in \mathcal{V}$ and $e_{ij} \in \{1..E\}$ is the label of the edge $(i, j) \in \mathcal{E}$. 

This graph can be represented as a sequence $s$ of vectors $V_k$ that can represent either a node or an edge. This sequence is constructed sequentially by adding each node label $x_i$ followed by all the edge labels between this node and the previously added ones $e_{ij}$:
\begin{equation*}
    s = \left(V_k\right)_{k=1..T} = \left(x_0, x_1, e_{01}, x_2, e_{02}, e_{12}, x_3, ...\right)
\end{equation*}
where $T$ is the length of the sequence. 
For practical reasons, we define all sequences to end with a particular node type vector acting as an \emph{end sequence} token. 
As an illustration, the representation of propane, ie a chain of three carbons with single bonds, could be $s=$ (\emph{carbon, carbon, single bond, carbon, no edge, single bond, end token}). Any sequence $s_k = \left(V_u\right)_{u=1..k-1}$ defines a valid subgraph $G_{s_k}$. 

Different sequences can represent the same molecule. Typically, the choice of the node indices $\mathcal{V}$ is arbitrary but changing the node indices will change the sequence s. We note $seqs(G)$ the ensemble of sequences allowed by our algorithm that represent graph $G$.

From a distribution over graphs $\mathcal{Q}(G)$ we can then trivially obtain a distribution on sequences $\mathcal{Q}(s)=\mathbb{E}_{G\sim\mathcal{Q}(G)}\mathcal{Q}(s|G)$ where $\mathcal{Q}(s|G)$ is uniform over $seqs(G)$. Notice that to any sequence corresponds only one graph, so that ${\mathcal{Q}(s) = \frac{\mathcal{Q}(G)}{|seqs(G)|}}$.

\subsubsection{Auto-regressive model}
\label{section: nn}

A learned distribution over sequences $\mathcal{P}(s)$ is then modelled in an auto-regressive fashion:

\begin{equation}
    \mathcal{P}(s=\left(V_k\right)_{k=1..T}) = \prod_k \mathcal{P}_\theta(V_k|G_{s_k})
\end{equation}
where $\mathcal{P}_\theta$ is a graph neural network defined below, with parameter $\theta$.

As there is only one graph per sequence, the corresponding distribution on graphs is simply $\mathcal{P}(G) = \sum_{s\in seqs(G)} \mathcal{P}(s)$.

We explicitly model $\mathcal{P}_\theta(V_k|G_{s_k})$ as a categorical distribution whose probability vector is the output of a graph neural network. 
First, node embeddings $h_i$ are produced by $L$ layers of convolution. As in \citet{shi2020graphaf}, we use an edge sensitive version of the standard GCN convolution:

\begin{equation}
\begin{aligned}
\label{eq:conv}
    h_i^{(l)} &= \sum_{e=1}^E ReLU\left(W_e h_i^{(l-1)} + V_e\sum_{j \in \mathcal{N}_e(i)} \frac{h_j^{(l)}}{\sqrt{d_i^e d_j^e}} + U_e\right)
\end{aligned}
\end{equation}
where $U_e$, $V_e$ and $W_e$ are learnable weight matrices, $\mathcal{N}_e(i) = \left\{j / (i,j)\in E \text{ and } e_{ij} = e\right\}$ is the set of neighbours of $i$ via an edge labelled $e$ and $d^e_i=|\mathcal{N}_e(i)|$ is the degree of node $i$ in the graph containing only edges labelled $e$. 
We set $h_i^{(0)}$ as the one hot embedding of $x_i$.

A global embedding is then defined by global addition pooling, ie $H = \sum_i h_i^{(l)}$.

Finally, two independent MLPs with \emph{softmax} activation are used to produce probability vectors for node and edge distributions:

\begin{equation}
\label{eq:mlp}
\begin{aligned}
\mathcal{P}(V_k=x_i|G_k) &= Cat(MLP^{node}(H)) \\
\mathcal{P}(V_k=e_{ij}|G_k) &= Cat(MLP^{edge}(H, h_i^{(l)}, h_j^{(l)}))
\end{aligned}
\end{equation}
where $Cat(.)$ is the categorical distribution with probabilities defined by the outputs of the network.

The network is then trained by minimizing the cross-entropy loss of each decision over the dataset.

\subsubsection{Further improvements and modifications}


\nlparagraph{Conditioning the network on molecule size}
\label{section:size}
In the next sections, we use among others a version of the model conditioned on molecule size \footnote{Additional care must be taken in training such network, as detailed in supplementary materials}. 
In order to do so, we redefine the convolution step in eq. \ref{eq:conv} as:

\begin{equation}
    h_i^{(l)} = \widehat{h}_i^{(l)} + C * \frac{s}{S}
\end{equation}
where $s$ is the final molecule size, $S$ is the maximal allowed molecule size, $C^{(l)}$ is a learned vector, and $\widehat{h}_i^{(l)}$ is the original definition in eq. \ref{eq:conv}.


\nlparagraph{Limiting the length of sequences}
\label{section:seq}
The length of the sequences is the major limitation of auto-regressive methods in term of time efficiency.
Moreover, most $V_k$ correspond to $e_{ij}=\text{\emph{no edge}}$, which seems like a waste of resources. 
A common approach is to always present the nodes ordered as a breadth first search(BFS) traversal of the graph, without repetition and rooted randomly. 
In that setting, there exists an upper limit $B \ge |i-j|$ depending on the maximal width of the BFS tree in the dataset. 
In the ZINC dataset, $B=12$. We can then omit all $e_{ij}$ if $|i-j|> B$ in $s$.
Notice that the same advantage occurs if the nodes are instead presented in the reversed order of a BFS exploration, meaning that the last explored node is presented first in the sequence. 
As opposed to the former case, in the latter, the last generated node is uniformly sampled among all the nodes. 

These two choices correspond to two definitions of $seqs(G)$. The difference between them will be exploited in section \ref{section:mcmc}.

\nlparagraph{Labelling the nodes for better cycle detection}
A common issue in graph generation is the occurrence of unrealistic cycles, never seen in the dataset. 
We believe this is due to the fact that atoms are anonymous, so that two atoms in a similar environment will have similar embeddings. Hence, the atom embedding cannot capture information about the distance to another specific atom. 
In order to partially solve the problem, we add a random label to each node to enable the model to distinguish between close pair of nodes, that have interacted with each other through convolutions, and far-away pairs. 
We implement that by redefining the initial node embedding: $h_i^{(0)} = (F(x_i),  \epsilon)$ where $F(x_i)$ is the one hot embedding of $x_i$ and $\epsilon$ is a random normal vector of fixed dimension. 
We found empirically that, without completely solving the problem, this improved the distribution of cycles in generated molecules.

\nlparagraph{Enforcing validity}
\label{section:validity}
A great advantage of auto-regressive models is that we can easily ensure that the maximal valency of each atom is respected. 
In order to do so, we simply set the probability of edges that would break the valency constraint to zero during generation. 
If the nodes are presented in standard BFS order, we can also ensure that the molecule remains connected during the generation process, by stopping the generation if a new node has no real bond to the rest of the molecule.

\subsection{Direct optimization in molecule manifold}
\label{section:mcmc}
In order to optimize molecule properties we use the Metropolis Hastings algorithm, with a slight adaptation in order to bias the simulated distribution, centered around the criterion to optimize, toward the learned distribution, which defines implicitly the criterion of realism. We refer to that method as Learned Realism Sampling (LRS).

\subsubsection{Algorithm}
First, we define a conditional transition probability $\mathcal{T}(G_1|G_0)$ from graph $G_0$ to graph $G_1$. 
Intuitively, $\mathcal{T}$ is a distribution over the neighborhood of $G_0$ in the ensemble of all graphs, which is not defined. 
Moreover, in order to ensure the realism of the proposal $G_1$, we need to sample from the learned distribution $\mathcal{P}$. 
In order to satisfy both conditions, we simply resample the last steps of the generative algorithm. 
More precisely, we start by sampling uniformly a generation sequence $s \in seqs(G_0)$ of length $M$, and remove the last elements corresponding to $\delta n\in\mathbb{N}$ node additions, yielding a shorter sequence that we call $s*$. We then rerun the generative model, starting from the intermediate graph $G_{s*}$. 

Then, given a score function $sc$ and temperature $T$, we define an acceptance probability $\mathcal{R}$ as:

\begin{equation}
\begin{aligned}
\mathcal{R}(G_1|G_0) &= \min\left(\exp\left(\frac{sc(G_1) - sc(G_0)}{T}\right), 1\right)
\end{aligned}
\label{eq:acceptance}
\end{equation}

From those two distributions and an initial graph $G_0$, we iteratively produce new graphs similarly to the Metropolis Hastings algorithm, as summed up in algorithm \ref{algo}:

\begin{algorithm}[h]
\begin{enumerate}
\item Sample a proposal graph $G^{proposal}_{n+1} \sim \mathcal{T}(G_{n+1}|G_n)$
\item Accept the proposal graph with probability $\mathcal{R}(G^{proposal}_{n+1}, G_n)$
\item If the proposal is accepted, set $G_{n+1} = G^{proposal}_{n+1}$. Else, set $G_{n+1}=G_n$.
\end{enumerate}
\caption{Learned Realism Sampling}
\label{algo}
\end{algorithm}

Notice that this is \emph{not} Metropolis Hastings algorithm as $\mathcal{T}$ is not symmetric, which should require a correction in $r$. Hence the simulated distribution is \emph{not} proportional to $F$. 

\begin{lemma}\ \\
Given a joint distribution $\mathcal{P}(s, G)$ over $\mathcal{S}\times\mathcal{G}$, if the conditional distribution $\mathcal{P}(s|G)$ is the uniform distribution over $seqs(G)$, then algorithm \ref{algo} is equivalent to Metropolis Hastings sampling for a distribution $\pi(G) \propto F(G)\mathcal{P}(G)$.
\label{lemma}
\end{lemma}
The proof is given in supplementary materials. 
Notice that the condition is fulfilled by the dataset distribution $\mathcal{Q}$, as detailed in \ref{section:sequence definition}. The result is only exact in the limit where the learned distribution $\mathcal{P}$ perfectly matches $\mathcal{Q}$.
This lemma should however describe the general tendency of the algorithm even if the approximation is imperfect, as there is no consistent bias during training. 

As a result, algorithm \ref{algo} produces graphs which both have high scores and are likely with regards to the learned distribution $\mathcal{P}$. 
The temperature parameter can be used to tune the importance of both those aspects or to perform annealing.

A great advantage of the method is that the learned distribution  is completely independent from the score calculation. 
Typically, $\mathcal{P}$ can be learned on large public datasets, like ZINC, while the score can be learned on scarce data, using very simple models, or even on no data at all if an analytic solution exists. 
Notice that many models  optimize the score by learning, in particular reinforcement learning ones. There is then a risk that some specificity of the score is learned at the expense of neglecting the dataset statistics. This is not the case of our model as the learning phase is only performed prior to optimization.

\subsubsection{Further improvements and variations}

\nlparagraph{Controlling the sampled graph size}
If the initial graph $G_0$ of size $N$ has few nodes compared to the median size of the training dataset, then the graph $G_1\sim\mathcal{T}(G_1|G_0)$ is likely to be much larger than $G_0$, breaking the intuition that $G_1$ should be similar to $G_0$. 
To correct that effect, a model conditioned on size can be used. We remove $\delta n_r$ nodes and condition the output size to be $N - \delta n_r + \delta n_a$. We sample both quantities $\delta n_r$ and $\delta n_a$ from a Poisson distribution with parameter $\lambda$.

\nlparagraph{Sampling the modified region uniformly}
As detailed in section \ref{section:seq}, the nodes can be presented in different orders. 
Following \citet{shi2020graphaf}, our basic model presents the nodes in a BFS order to improve efficiency. The removed atoms are the ones visited last. Thus, the probability to be removed is not the same among all nodes. For example, in a long carbon chain, the atoms in the middle will almost never be removed.

In order to circumvent that effect and without loss of efficiency, the nodes can be presented in reversed BFS order, so that the root is presented last. 
By doing so, the first removed node is sampled uniformly among the graph. 
Notice that if $\delta n_r > 1$, this modification is not enough to ensure that nodes have a uniform probability to be removed, as more connected nodes will be selected more frequently. 
Still, we find that this correction is enough for substantial improvement while remaining very simple.

\section{Experiments}
We compare our model to multiple state-of-the-art models described in more details in section \ref{related}.

\subsection{Distribution Learning}

\begin{table}[t]
\caption{Distribution learning performance of all the baselines. Validity(Val.), Novelty(Nov.), Unicity(Uni.), Kl, FCD scores are presented. ($\uparrow$) and ($\downarrow$) indicates respectively that higher or lower is better. VJTNN and MIMOSA are missing from this table, as they cannot be trained only for distribution learning. All models have been retrained using the official released code, except for MRNN which is not available. }
\label{results:distribution learning}
\vskip 0.15in
\centering
\begin{tabularx}{\linewidth}{@{}lXXXXX@{}}
\toprule
~& Val.($\uparrow$) & Nov.($\uparrow$) & Uni.($\uparrow$) & KL($\downarrow$) & FCD($\downarrow$)\\
 \midrule
JTVAE & 1.00 & 1.0 & 1.000 & 0.078 & 1.46\\
GCPN &  0.20 & 1.0 & 1.000 & 0.604 & 12.49\\
MRNN & 0.65 & 1.0 & 0.999 & - & -\\
GraphAF & 0.68 &1.0 & 0.991 & 0.2294 &10.06\\
SMILES LSTM & 0.87 & 0.999 & 1.000 & 0.178 & 0.515\\
 \midrule
AR & 0.97 & 1.0 & 1.000 & 0.060 & 1.48\\
AR-SC & 0.97& 1.0 & 1.000 & 0.058 & 1.50\\
AR-RBFS & 0.88 & 1.0 & 1.000 & 0.102 & 2.83\\
 \bottomrule
\end{tabularx}
\end{table}

\newcommand{\svgsw}{\textwidth}
\newcommand{\svgratio}{0.35}
\newcommand{\svginvratio}{3}

\begin{figure*}[htb]
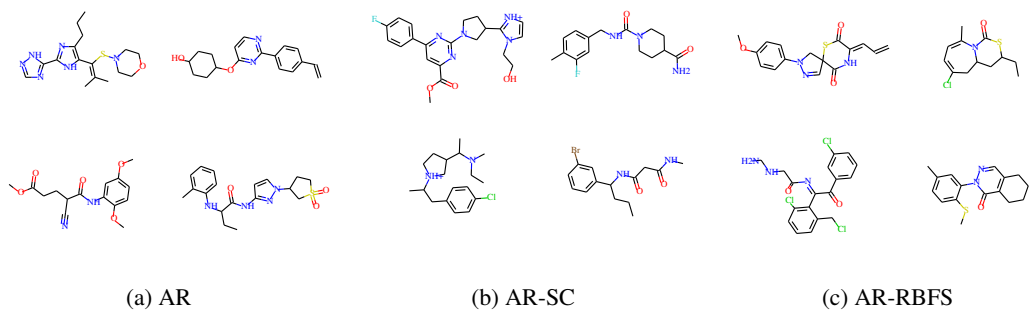

\centering{
\begin{subfigure}[b]{0.3\textwidth}
\def\svgwidth{\svginvratio\svgsw}
\scalebox{\svgratio}{\input{basic_2x2_compressed.pdf_tex}}
\caption{AR}
\end{subfigure}
\hfill
\begin{subfigure}[b]{0.3\textwidth}
\def\svgwidth{\svginvratio\svgsw}
\scalebox{\svgratio}{\input{sc_2x2_compressed.pdf_tex}}
\caption{AR-SC}
\end{subfigure}
\hfill
\begin{subfigure}[b]{0.3\textwidth}
\def\svgwidth{\svginvratio\svgsw}
\scalebox{\svgratio}{\input{rbfs_2x2_compressed.pdf_tex}}
\caption{AR-RBFS}
\end{subfigure}
}
\caption{Samples from the different versions of our model. No cherrypicking.}
\label{figure:generation}
\end{figure*}

Our method begins by training the autoregressive models described above to learn the dataset distribution. 
This is obviously a critical step, as the subsequent algorithm will produce samples biased toward that distribution. 

We train three versions of our autoregressive model:
\begin{itemize}
\item The basic model described in \ref{section:armodel}, simply referred to as AR.
\item A model conditioned on size as described in section \ref{section:size}, referred to as AR-SC
\item A model conditioned on size, and with nodes presented in reversed BFS order, as in section \ref{section:seq}, referred to as AR-RBFS
\end{itemize}
The model is trained on the ZINC250K dataset from \cite{irwin2005zinc}.
Implementation details can be found in supplementary materials, such as input format and hyperparameters.

The model is compared to state-of-the-art baselines using multiple metrics, all detailed in appendix \ref{metrics definition}.
We use the standard validity, novelty and uniqueness metrics, measured for a set of 10,000 molecules. 
Following  \citet{brown2019guacamol}, we also provide two other metrics: the Kullback-Leibler divergence between the distributions of the generated molecules and the dataset, evaluated on many molecular descriptors. The Fréchet ChemNet Distance (FCD) \citep{preuer2018frechet} is a molecular version of the Fréchet Inception Distance. 
Finally, we provide images of generated molecules in figure \ref{figure:generation}.


The results are displayed in table \ref{results:distribution learning}, showing that our model achieves state-of-the-art performance, and generally better than reinforcement learning based methods. 
The RBFS model achieves slightly lower validity, KL and FCD, which is accountable to its more complex task as it must also ensure molecule connectivity.

\subsection{Score optimization}
We then use the three models from the previous section to perform our proposed LRS methods. Our method using the AR-X model is referred to as LRS-X. In the first section, we demonstrate that optimization can produce irrealistic molecules. In the second section, we show that although our model has the additional limitation to produce realistic molecules, it can still perform as well or better than state-of-the-art methods in the task of molecule improvement. 

\subsubsection{Unsafe optimization can lead to generation of unrealistic molecules}
\label{section: unconstrained optimisation}
\newcommand{\molsize}{8cm}
\newcommand{\molratio}{0.25}
\newcommand{\rw}{0.2}
\begin{figure*}[htb]
\begin{subtable}[b]{\textwidth}
\begin{tabularx}{\textwidth}{@{}cXccXc@{}}
 (4.52) & & GCPN (7.98) & JTVAE (5.30) & & LRS (5.42)\\
\scalebox{\molratio}{\def\svgwidth{\molsize}
\begingroup%
  \makeatletter%
  \providecommand\color[2][]{%
    \errmessage{(Inkscape) Color is used for the text in Inkscape, but the package 'color.sty' is not loaded}%
    \renewcommand\color[2][]{}%
  }%
  \providecommand\transparent[1]{%
    \errmessage{(Inkscape) Transparency is used (non-zero) for the text in Inkscape, but the package 'transparent.sty' is not loaded}%
    \renewcommand\transparent[1]{}%
  }%
  \providecommand\rotatebox[2]{#2}%
  \newcommand*\fsize{\dimexpr\f@size pt\relax}%
  \newcommand*\lineheight[1]{\fontsize{\fsize}{#1\fsize}\selectfont}%
  \ifx\svgwidth\undefined%
    \setlength{\unitlength}{300bp}%
    \ifx\svgscale\undefined%
      \relax%
    \else%
      \setlength{\unitlength}{\unitlength * \real{\svgscale}}%
    \fi%
  \else%
    \setlength{\unitlength}{\svgwidth}%
  \fi%
  \global\let\svgwidth\undefined%
  \global\let\svgscale\undefined%
  \makeatother%
  \begin{picture}(1,1)%
    \lineheight{1}%
    \setlength\tabcolsep{0pt}%
    \put(0,0){\includegraphics[width=\unitlength,page=1]{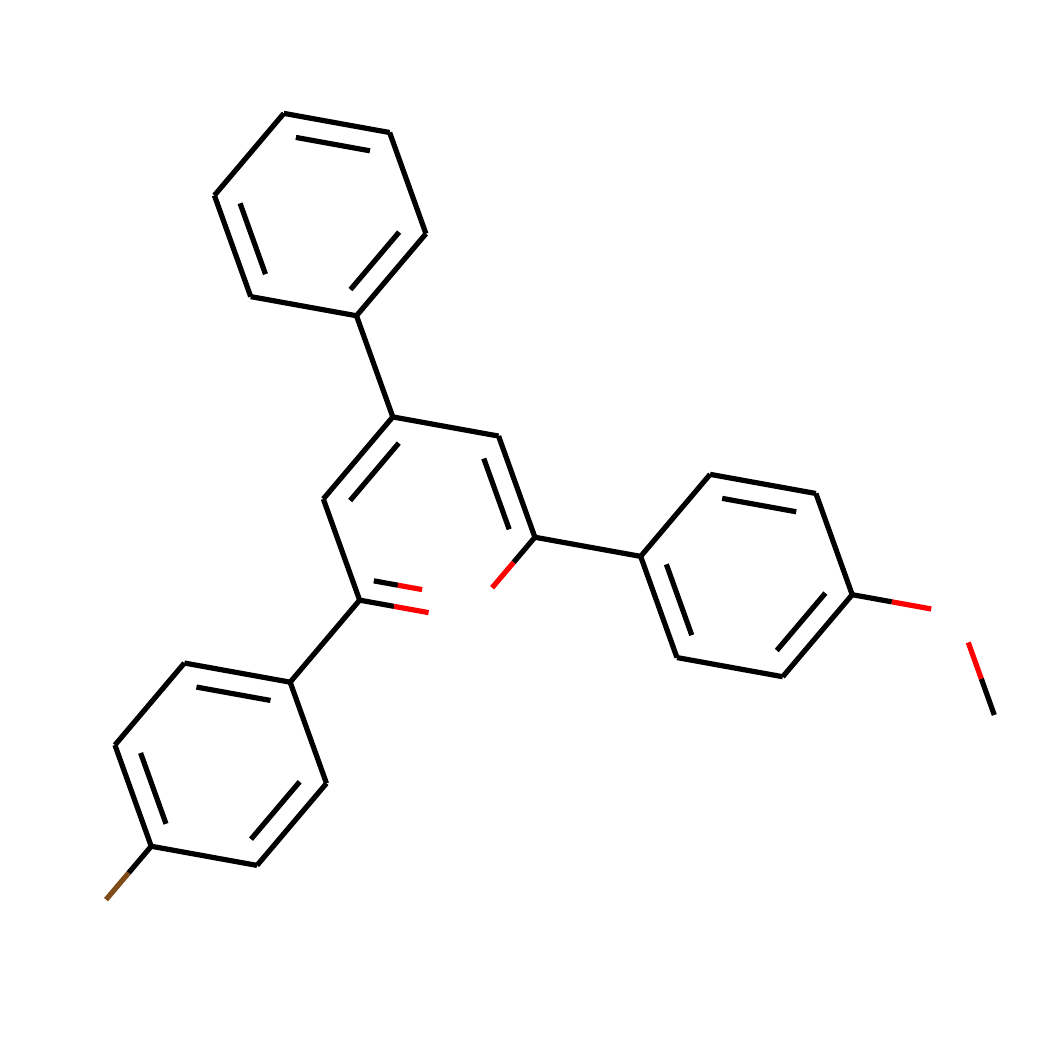}}%
    \put(0.89398247,0.38321499){\color[rgb]{1,0,0}\makebox(0,0)[lt]{\lineheight{1.25}\smash{\begin{tabular}[t]{l}O\end{tabular}}}}%
    \put(0.0454545,0.08132751){\color[rgb]{0.49803922,0.29803922,0.09803922}\makebox(0,0)[lt]{\lineheight{1.25}\smash{\begin{tabular}[t]{l}Br\end{tabular}}}}%
    \put(0.41145,0.37521751){\color[rgb]{1,0,0}\makebox(0,0)[lt]{\lineheight{1.25}\smash{\begin{tabular}[t]{l}O+\end{tabular}}}}%
  \end{picture}%
\endgroup%
} & & \scalebox{\molratio}{\def\svgwidth{\molsize}\input{GCPN_0.pdf_tex}} & \scalebox{\molratio}{\def\svgwidth{\molsize}
\begingroup%
  \makeatletter%
  \providecommand\color[2][]{%
    \errmessage{(Inkscape) Color is used for the text in Inkscape, but the package 'color.sty' is not loaded}%
    \renewcommand\color[2][]{}%
  }%
  \providecommand\transparent[1]{%
    \errmessage{(Inkscape) Transparency is used (non-zero) for the text in Inkscape, but the package 'transparent.sty' is not loaded}%
    \renewcommand\transparent[1]{}%
  }%
  \providecommand\rotatebox[2]{#2}%
  \newcommand*\fsize{\dimexpr\f@size pt\relax}%
  \newcommand*\lineheight[1]{\fontsize{\fsize}{#1\fsize}\selectfont}%
  \ifx\svgwidth\undefined%
    \setlength{\unitlength}{300bp}%
    \ifx\svgscale\undefined%
      \relax%
    \else%
      \setlength{\unitlength}{\unitlength * \real{\svgscale}}%
    \fi%
  \else%
    \setlength{\unitlength}{\svgwidth}%
  \fi%
  \global\let\svgwidth\undefined%
  \global\let\svgscale\undefined%
  \makeatother%
  \begin{picture}(1,1)%
    \lineheight{1}%
    \setlength\tabcolsep{0pt}%
    \put(0,0){\includegraphics[width=\unitlength,page=1]{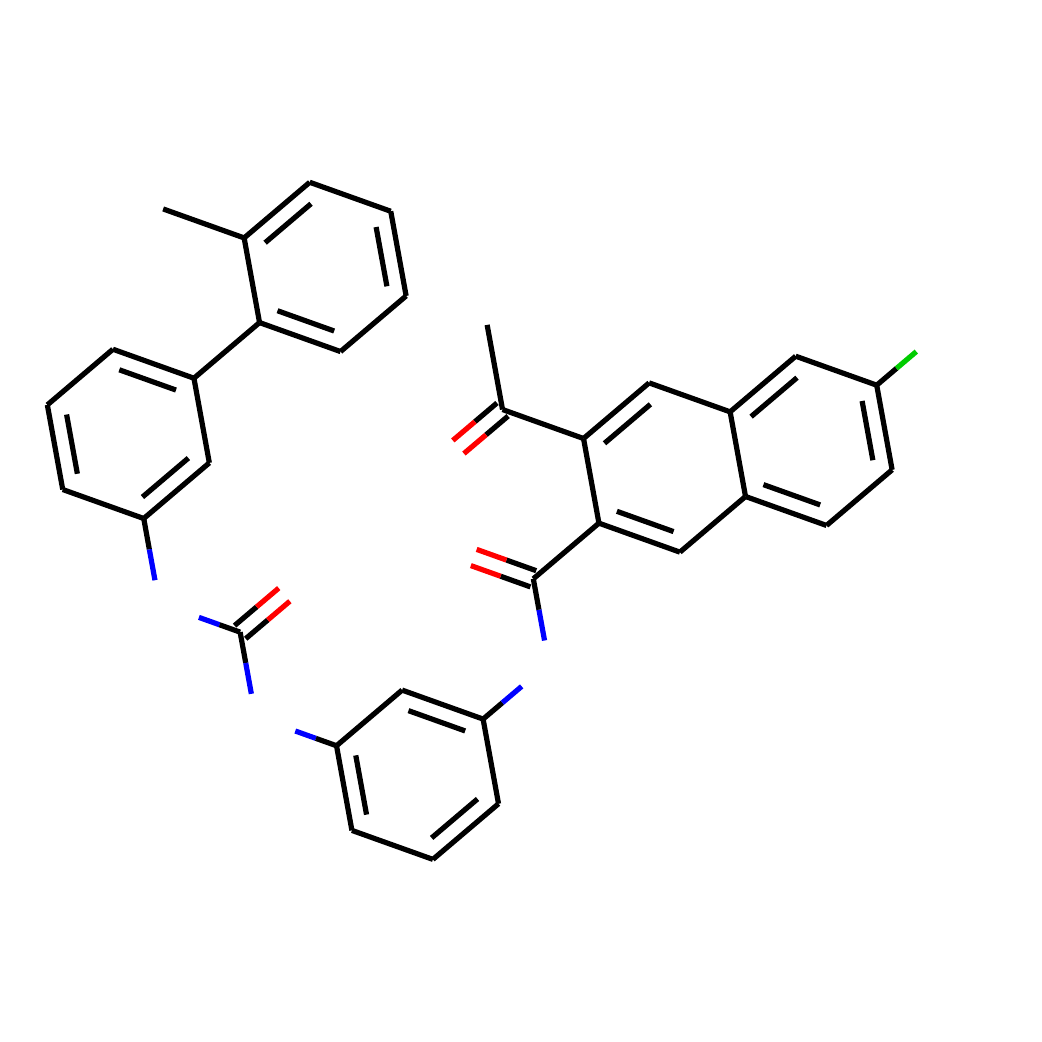}}%
    \put(0.114594,0.39895248){\color[rgb]{0,0,1}\makebox(0,0)[lt]{\lineheight{1.25}\smash{\begin{tabular}[t]{l}NH\end{tabular}}}}%
    \put(0.27293501,0.4245525){\color[rgb]{1,0,0}\makebox(0,0)[lt]{\lineheight{1.25}\smash{\begin{tabular}[t]{l}O\end{tabular}}}}%
    \put(0.20713024,0.28987251){\color[rgb]{0,0,1}\makebox(0,0)[lt]{\lineheight{1.25}\smash{\begin{tabular}[t]{l}NH\end{tabular}}}}%
    \put(0.48859501,0.34107246){\color[rgb]{0,0,1}\makebox(0,0)[lt]{\lineheight{1.25}\smash{\begin{tabular}[t]{l}NH\end{tabular}}}}%
    \put(0.41366749,0.45014999){\color[rgb]{1,0,0}\makebox(0,0)[lt]{\lineheight{1.25}\smash{\begin{tabular}[t]{l}O\end{tabular}}}}%
    \put(0.87967247,0.6616275){\color[rgb]{0,0.8,0}\makebox(0,0)[lt]{\lineheight{1.25}\smash{\begin{tabular}[t]{l}Cl\end{tabular}}}}%
    \put(0.39888748,0.53140251){\color[rgb]{1,0,0}\makebox(0,0)[lt]{\lineheight{1.25}\smash{\begin{tabular}[t]{l}O\end{tabular}}}}%
  \end{picture}%
\endgroup%
} & &  \scalebox{\molratio}{\def\svgwidth{\molsize}
\begingroup%
  \makeatletter%
  \providecommand\color[2][]{%
    \errmessage{(Inkscape) Color is used for the text in Inkscape, but the package 'color.sty' is not loaded}%
    \renewcommand\color[2][]{}%
  }%
  \providecommand\transparent[1]{%
    \errmessage{(Inkscape) Transparency is used (non-zero) for the text in Inkscape, but the package 'transparent.sty' is not loaded}%
    \renewcommand\transparent[1]{}%
  }%
  \providecommand\rotatebox[2]{#2}%
  \newcommand*\fsize{\dimexpr\f@size pt\relax}%
  \newcommand*\lineheight[1]{\fontsize{\fsize}{#1\fsize}\selectfont}%
  \ifx\svgwidth\undefined%
    \setlength{\unitlength}{300bp}%
    \ifx\svgscale\undefined%
      \relax%
    \else%
      \setlength{\unitlength}{\unitlength * \real{\svgscale}}%
    \fi%
  \else%
    \setlength{\unitlength}{\svgwidth}%
  \fi%
  \global\let\svgwidth\undefined%
  \global\let\svgscale\undefined%
  \makeatother%
  \begin{picture}(1,1)%
    \lineheight{1}%
    \setlength\tabcolsep{0pt}%
    \put(0,0){\includegraphics[width=\unitlength,page=1]{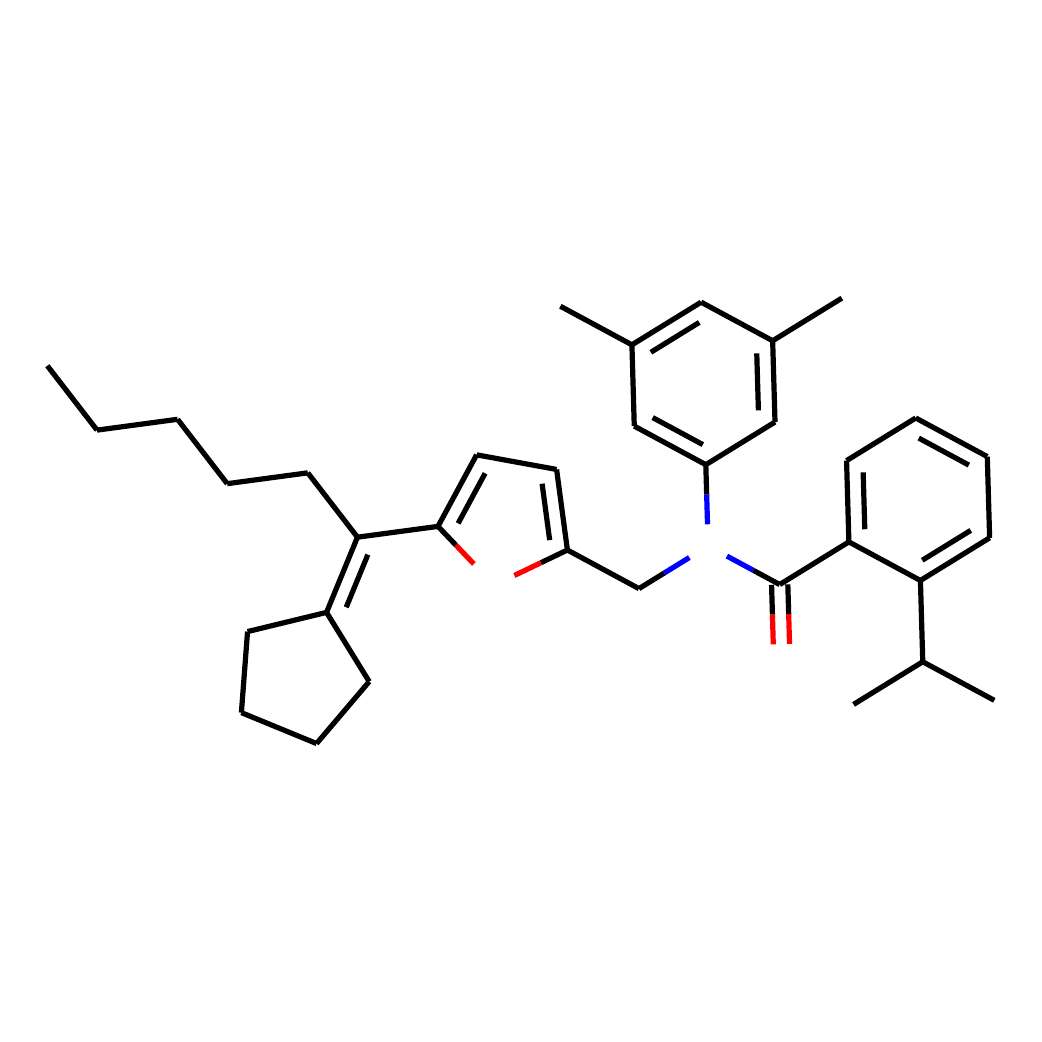}}%
    \put(0.66178749,0.45507252){\color[rgb]{0,0,1}\makebox(0,0)[lt]{\lineheight{1.25}\smash{\begin{tabular}[t]{l}N\end{tabular}}}}%
    \put(0.73131752,0.34000999){\color[rgb]{1,0,0}\makebox(0,0)[lt]{\lineheight{1.25}\smash{\begin{tabular}[t]{l}O\end{tabular}}}}%
    \put(0.45493,0.4173975){\color[rgb]{1,0,0}\makebox(0,0)[lt]{\lineheight{1.25}\smash{\begin{tabular}[t]{l}O\end{tabular}}}}%
  \end{picture}%
\endgroup%
}\\
(4.23) & & GraphAF (12.23) & All SMILES (12.34) & & LRS-SC (5.90)\\

\scalebox{\molratio}{\def\svgwidth{\molsize}
\begingroup%
  \makeatletter%
  \providecommand\color[2][]{%
    \errmessage{(Inkscape) Color is used for the text in Inkscape, but the package 'color.sty' is not loaded}%
    \renewcommand\color[2][]{}%
  }%
  \providecommand\transparent[1]{%
    \errmessage{(Inkscape) Transparency is used (non-zero) for the text in Inkscape, but the package 'transparent.sty' is not loaded}%
    \renewcommand\transparent[1]{}%
  }%
  \providecommand\rotatebox[2]{#2}%
  \newcommand*\fsize{\dimexpr\f@size pt\relax}%
  \newcommand*\lineheight[1]{\fontsize{\fsize}{#1\fsize}\selectfont}%
  \ifx\svgwidth\undefined%
    \setlength{\unitlength}{300bp}%
    \ifx\svgscale\undefined%
      \relax%
    \else%
      \setlength{\unitlength}{\unitlength * \real{\svgscale}}%
    \fi%
  \else%
    \setlength{\unitlength}{\svgwidth}%
  \fi%
  \global\let\svgwidth\undefined%
  \global\let\svgscale\undefined%
  \makeatother%
  \begin{picture}(1,1)%
    \lineheight{1}%
    \setlength\tabcolsep{0pt}%
    \put(0,0){\includegraphics[width=\unitlength,page=1]{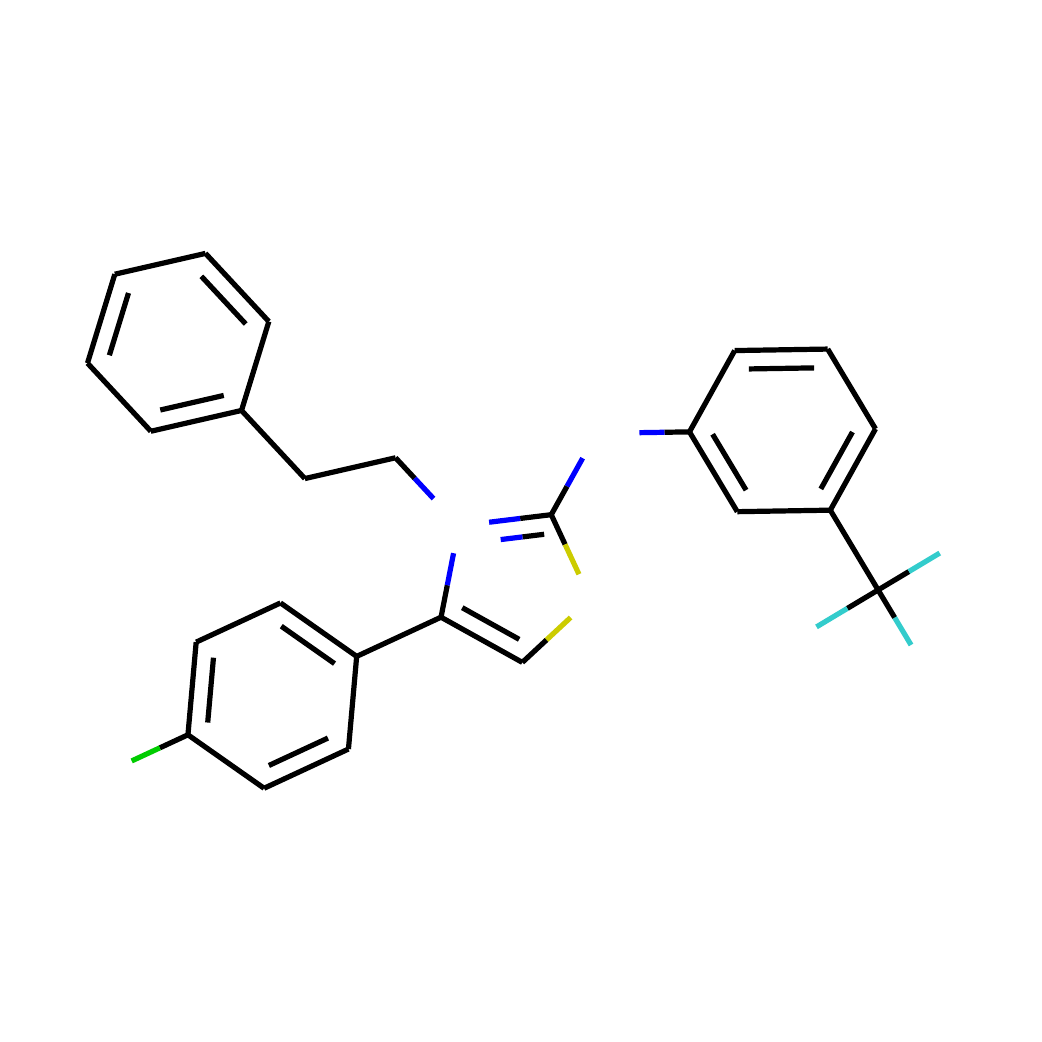}}%
    \put(0.8714225,0.33319252){\color[rgb]{0.2,0.8,0.8}\makebox(0,0)[lt]{\lineheight{1.25}\smash{\begin{tabular}[t]{l}F\end{tabular}}}}%
    \put(0.90213997,0.4557){\color[rgb]{0.2,0.8,0.8}\makebox(0,0)[lt]{\lineheight{1.25}\smash{\begin{tabular}[t]{l}F\end{tabular}}}}%
    \put(0.74891502,0.36390999){\color[rgb]{0.2,0.8,0.8}\makebox(0,0)[lt]{\lineheight{1.25}\smash{\begin{tabular}[t]{l}F\end{tabular}}}}%
    \put(0.531325,0.56014999){\color[rgb]{0,0,1}\makebox(0,0)[lt]{\lineheight{1.25}\smash{\begin{tabular}[t]{l}NH\end{tabular}}}}%
    \put(0.547775,0.401105){\color[rgb]{0.8,0.8,0}\makebox(0,0)[lt]{\lineheight{1.25}\smash{\begin{tabular}[t]{l}S\end{tabular}}}}%
    \put(0.0724435,0.23314003){\color[rgb]{0,0.8,0}\makebox(0,0)[lt]{\lineheight{1.25}\smash{\begin{tabular}[t]{l}Cl\end{tabular}}}}%
    \put(0.4115625,0.46894749){\color[rgb]{0,0,1}\makebox(0,0)[lt]{\lineheight{1.25}\smash{\begin{tabular}[t]{l}N+\end{tabular}}}}%
  \end{picture}%
\endgroup%
} & & \scalebox{\molratio}{\def\svgwidth{\molsize}\input{GraphAF_0.pdf_tex}} & \scalebox{\molratio}{\def\svgwidth{\molsize}\input{all_smiles_0.pdf_tex}} & & \scalebox{\molratio}{\def\svgwidth{\molsize}\input{ours_size_1.pdf_tex}}\\
(4.22) & & MRNN (8.63) & SMILES LSTM (11.87) & & LRS-RBFS (6.98)\\

\scalebox{\molratio}{\def\svgwidth{\molsize}
\begingroup%
  \makeatletter%
  \providecommand\color[2][]{%
    \errmessage{(Inkscape) Color is used for the text in Inkscape, but the package 'color.sty' is not loaded}%
    \renewcommand\color[2][]{}%
  }%
  \providecommand\transparent[1]{%
    \errmessage{(Inkscape) Transparency is used (non-zero) for the text in Inkscape, but the package 'transparent.sty' is not loaded}%
    \renewcommand\transparent[1]{}%
  }%
  \providecommand\rotatebox[2]{#2}%
  \newcommand*\fsize{\dimexpr\f@size pt\relax}%
  \newcommand*\lineheight[1]{\fontsize{\fsize}{#1\fsize}\selectfont}%
  \ifx\svgwidth\undefined%
    \setlength{\unitlength}{300bp}%
    \ifx\svgscale\undefined%
      \relax%
    \else%
      \setlength{\unitlength}{\unitlength * \real{\svgscale}}%
    \fi%
  \else%
    \setlength{\unitlength}{\svgwidth}%
  \fi%
  \global\let\svgwidth\undefined%
  \global\let\svgscale\undefined%
  \makeatother%
  \begin{picture}(1,1)%
    \lineheight{1}%
    \setlength\tabcolsep{0pt}%
    \put(0,0){\includegraphics[width=\unitlength,page=1]{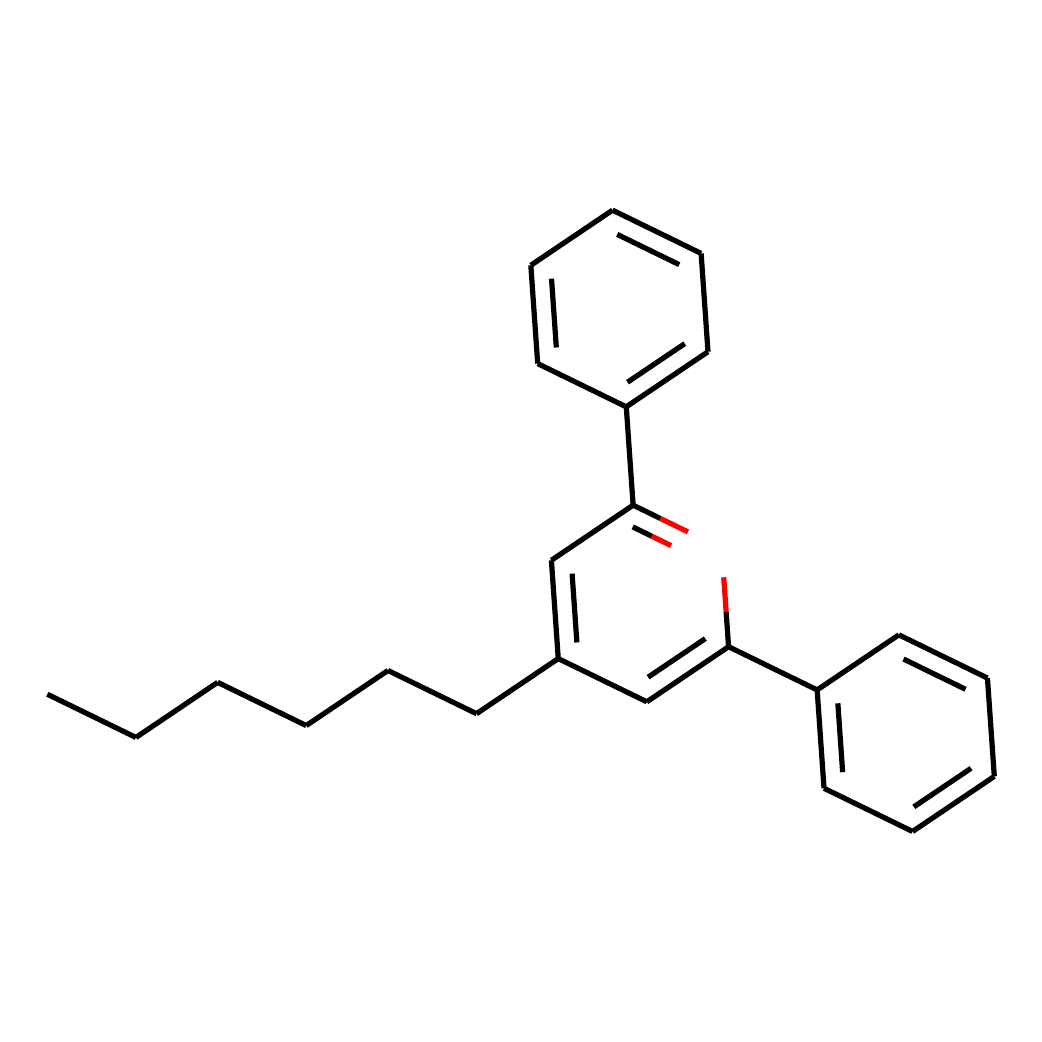}}%
    \put(0.66043251,0.44579498){\color[rgb]{1,0,0}\makebox(0,0)[lt]{\lineheight{1.25}\smash{\begin{tabular}[t]{l}O+\end{tabular}}}}%
  \end{picture}%
\endgroup%
} & & \scalebox{\molratio}{\def\svgwidth{\molsize}\input{MRNN_0.pdf_tex}} & \scalebox{\molratio}{\def\svgwidth{\molsize}
\begingroup%
  \makeatletter%
  \providecommand\color[2][]{%
    \errmessage{(Inkscape) Color is used for the text in Inkscape, but the package 'color.sty' is not loaded}%
    \renewcommand\color[2][]{}%
  }%
  \providecommand\transparent[1]{%
    \errmessage{(Inkscape) Transparency is used (non-zero) for the text in Inkscape, but the package 'transparent.sty' is not loaded}%
    \renewcommand\transparent[1]{}%
  }%
  \providecommand\rotatebox[2]{#2}%
  \newcommand*\fsize{\dimexpr\f@size pt\relax}%
  \newcommand*\lineheight[1]{\fontsize{\fsize}{#1\fsize}\selectfont}%
  \ifx\svgwidth\undefined%
    \setlength{\unitlength}{300bp}%
    \ifx\svgscale\undefined%
      \relax%
    \else%
      \setlength{\unitlength}{\unitlength * \real{\svgscale}}%
    \fi%
  \else%
    \setlength{\unitlength}{\svgwidth}%
  \fi%
  \global\let\svgwidth\undefined%
  \global\let\svgscale\undefined%
  \makeatother%
  \begin{picture}(1,1)%
    \lineheight{1}%
    \setlength\tabcolsep{0pt}%
    \put(0,0){\includegraphics[width=\unitlength,page=1]{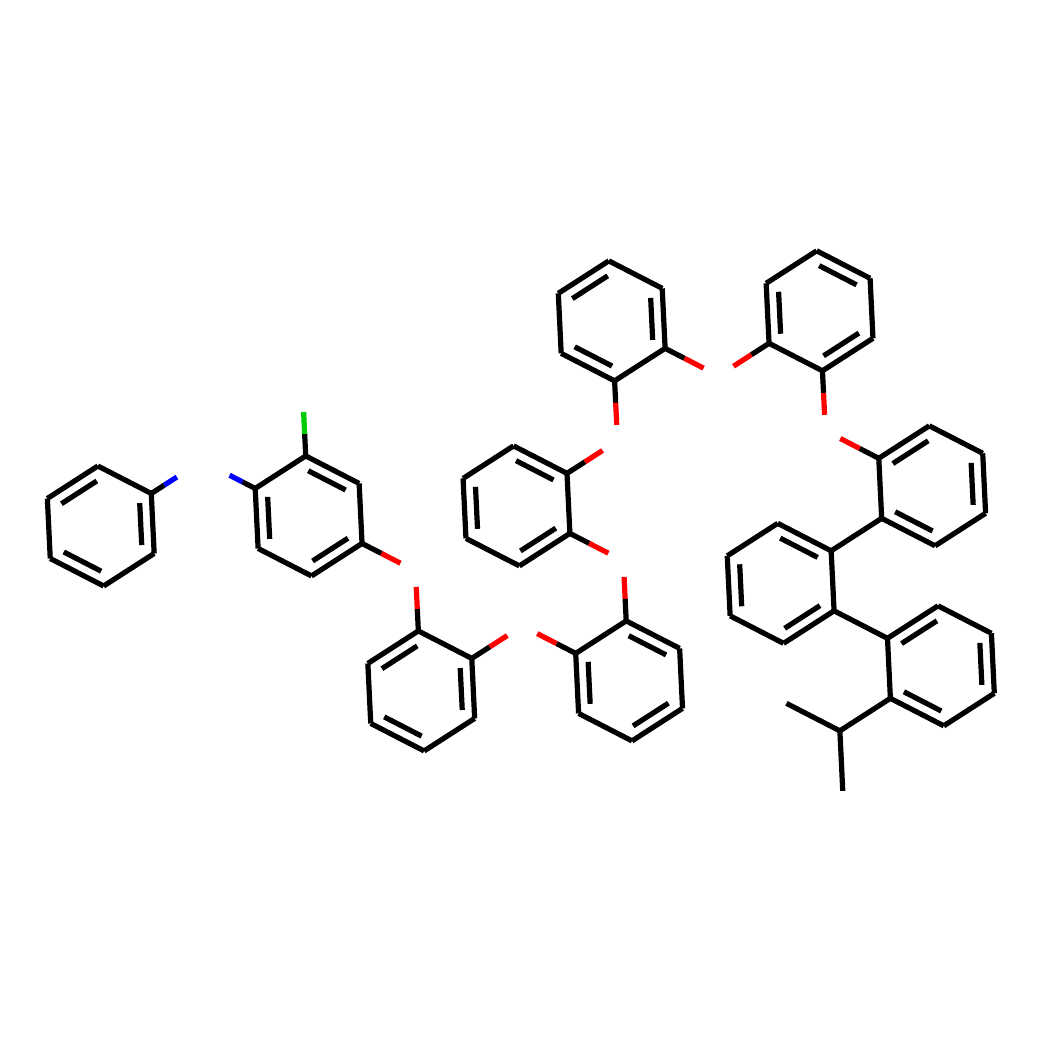}}%
    \put(0.77799248,0.57102749){\color[rgb]{1,0,0}\makebox(0,0)[lt]{\lineheight{1.25}\smash{\begin{tabular}[t]{l}O\end{tabular}}}}%
    \put(0.67543503,0.62380749){\color[rgb]{1,0,0}\makebox(0,0)[lt]{\lineheight{1.25}\smash{\begin{tabular}[t]{l}O\end{tabular}}}}%
    \put(0.57844749,0.56138){\color[rgb]{1,0,0}\makebox(0,0)[lt]{\lineheight{1.25}\smash{\begin{tabular}[t]{l}O\end{tabular}}}}%
    \put(0.58401749,0.44617249){\color[rgb]{1,0,0}\makebox(0,0)[lt]{\lineheight{1.25}\smash{\begin{tabular}[t]{l}O\end{tabular}}}}%
    \put(0.48702999,0.383745){\color[rgb]{1,0,0}\makebox(0,0)[lt]{\lineheight{1.25}\smash{\begin{tabular}[t]{l}O\end{tabular}}}}%
    \put(0.3844725,0.436525){\color[rgb]{1,0,0}\makebox(0,0)[lt]{\lineheight{1.25}\smash{\begin{tabular}[t]{l}O\end{tabular}}}}%
    \put(0.1670615,0.542085){\color[rgb]{0,0,1}\makebox(0,0)[lt]{\lineheight{1.25}\smash{\begin{tabular}[t]{l}NH\end{tabular}}}}%
    \put(0.27327999,0.60451248){\color[rgb]{0,0.8,0}\makebox(0,0)[lt]{\lineheight{1.25}\smash{\begin{tabular}[t]{l}Cl\end{tabular}}}}%
  \end{picture}%
\endgroup%
} & & \scalebox{\molratio}{\def\svgwidth{\molsize}
\begingroup%
  \makeatletter%
  \providecommand\color[2][]{%
    \errmessage{(Inkscape) Color is used for the text in Inkscape, but the package 'color.sty' is not loaded}%
    \renewcommand\color[2][]{}%
  }%
  \providecommand\transparent[1]{%
    \errmessage{(Inkscape) Transparency is used (non-zero) for the text in Inkscape, but the package 'transparent.sty' is not loaded}%
    \renewcommand\transparent[1]{}%
  }%
  \providecommand\rotatebox[2]{#2}%
  \newcommand*\fsize{\dimexpr\f@size pt\relax}%
  \newcommand*\lineheight[1]{\fontsize{\fsize}{#1\fsize}\selectfont}%
  \ifx\svgwidth\undefined%
    \setlength{\unitlength}{300bp}%
    \ifx\svgscale\undefined%
      \relax%
    \else%
      \setlength{\unitlength}{\unitlength * \real{\svgscale}}%
    \fi%
  \else%
    \setlength{\unitlength}{\svgwidth}%
  \fi%
  \global\let\svgwidth\undefined%
  \global\let\svgscale\undefined%
  \makeatother%
  \begin{picture}(1,1)%
    \lineheight{1}%
    \setlength\tabcolsep{0pt}%
    \put(0,0){\includegraphics[width=\unitlength,page=1]{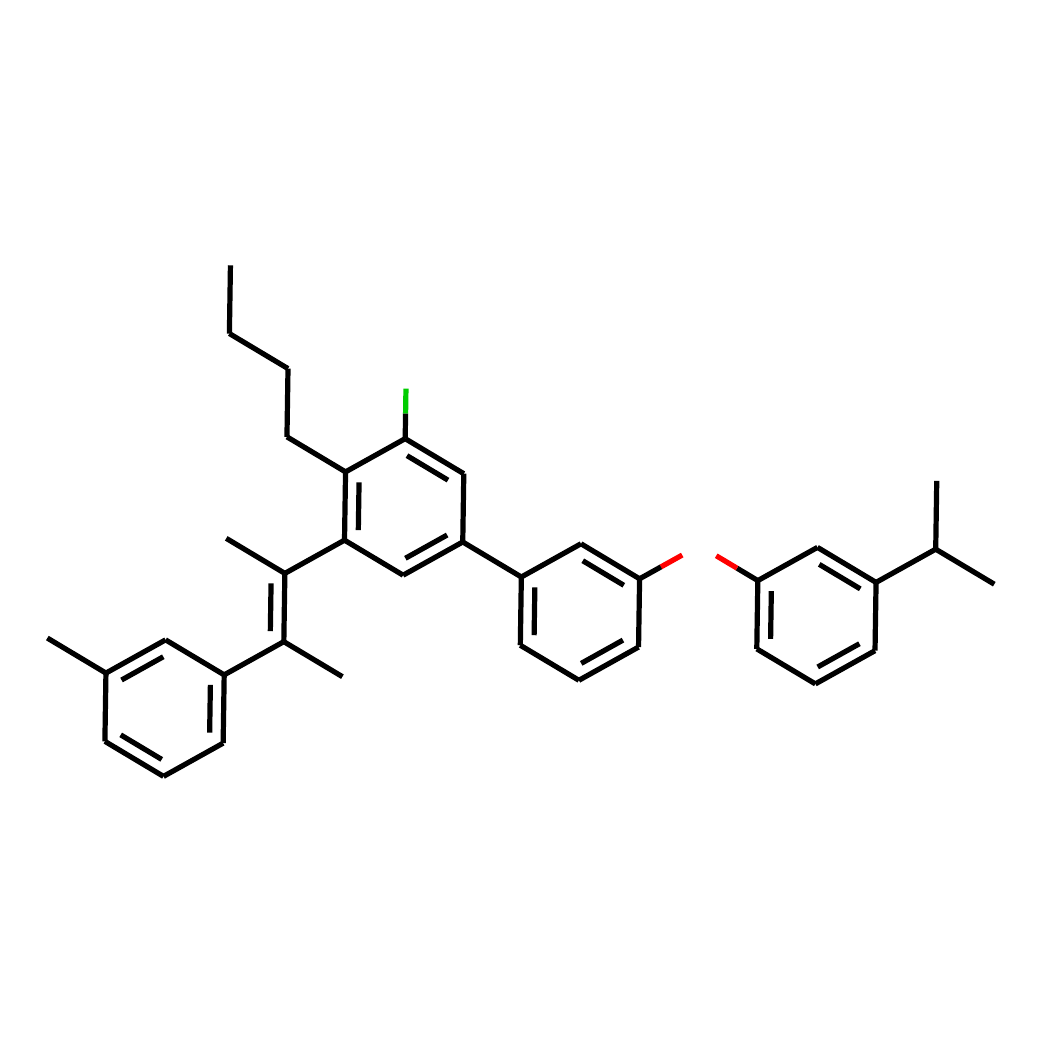}}%
    \put(0.65494247,0.45877251){\color[rgb]{1,0,0}\makebox(0,0)[lt]{\lineheight{1.25}\smash{\begin{tabular}[t]{l}O\end{tabular}}}}%
    \put(0.3701825,0.626945){\color[rgb]{0,0.8,0}\makebox(0,0)[lt]{\lineheight{1.25}\smash{\begin{tabular}[t]{l}Cl\end{tabular}}}}%
  \end{picture}%
\endgroup%
}\\
(a) Dataset molecules & & \multicolumn{2}{c}{(b) Baselines} & & (c) Ours \\

\end{tabularx}
\end{subtable}
\caption{Candidates with largest NPLogP from the dataset (a), or sampled from the baselines (b) and our model (c).}
\label{figure:high NPLogP}
\end{figure*}

A standard task is to generate molecules with the largest score possible. A common proof-of-principle score is the normalized penalized water octanol partition logP coefficient (NPlogP). This score happens to be easy to maximize, if no realism constraint is required. For example, a very long carbon chain, although unlikely in the training dataset, has a very high score. We show that many models will disregard the dataset statistics in order to optimize the score.

In general, it is hard to provide a metric, or even to evaluate, the realism of a molecule regarding a dataset. In this particular case, the molecules obtained from a model that disregarded its training set are particularly odd, so that we can attempt to quantify their quality. As in literature, we provide the top 3 molecules generated by each model. To estimate their realism, we provide their mean log-likelihood in feature space: First, we use a set of property descriptors as for the KL metric in previous section. Second, we use the ChemNet features. Details can be found in appendix. Results are grouped in table \ref{results:property optimization}. Finally, we provide images of the top candidate for each model in figure \ref{figure:high NPLogP}.

First, we notice that \emph{all} models produce NPlogP scores that are considerably above what can be found in the dataset, which is already an alarming sign. We also notice that most models produce extremely unlikely molecules. Indeed, as can be seen on figure \ref{figure:high NPLogP}, most models only learned to repeat the same pattern many times. On the other hand, both JTVAE and our models produce more reasonable candidates (with some doubt concerning the descriptor log likelihood of LRS-SC and LRS-RBFS). We conclude that our model is indeed efficiently able to maximize a criterion while better enforcing realism of the generated candidates.

\begin{table}[hb]
\caption{Top three performances of NPLogP optimization for all models and log-likelihood against descriptors and ChemNet features.}
\label{results:property optimization}
\vskip 0.15in
\centering
\begin{tabularx}{\linewidth}{@{}lXXXXX@{}}
 \toprule
 & \multicolumn{3}{c}{NPLogP ($\uparrow$)} & \multicolumn{2}{c}{Log-Likelihood ($\uparrow$)}\\ 
 \cmidrule(r){2-4} \cmidrule(l){5-6}
 & 1st & 2nd & 3rd & Descriptors & Chemnet Feat.\\
 \midrule
Dataset & 4.52 & 4.23 & 4.22 & -37 & 291\\
 \hline
 JTVAE & 5.30 & 4.93 & 4.49 & -38 & 283\\
 GCPN & 7.98 & 7.85 & 7.80 & -101 & -226 \\
 MRNN & 8.63 & 6.90 & 4.73 & -44 & -125\\
 GraphAF & 12.23 & 11.29 & 11.05 & -286 & 130\\
 All-SMILES & 12.34 & 12.16 & 12.05 & -210 & -333\\
 \small{SMILES} LSTM & 11.87 & 11.84 & 11.70 & -164 & -171\\
 \midrule
 LRS & 5.42  & 5.22 & 5.11 & -36  & 427\\
 LRS-SC & 5.90& 5.90 & 5.83 & -47 & 284\\
 LRS-RBFS & 6.98  & 6.92 & 6.69 & -51  & 293\\
 \bottomrule
\end{tabularx}
\end{table}


\subsubsection{Molecule improvement}

\begin{table}[t]
\caption{NPlogP score improvement for a given similarity threshold $\delta$. Results for JTVAE are taken from \citet{jin2018learning}.}
\label{results:molecule improvement NPlogP}
\vskip 0.15in
\centering
\begin{tabularx}{\linewidth}{@{}lXXX@{}}
 \toprule
 & \multicolumn{3}{c}{NPlogP}\\
 \cmidrule(l){2-4}
 & $\delta = 0.2$ & $\delta = 0.4$ & $\delta = 0.6$\\
 \midrule
 JTVAE & -  & $1.03 \pm 1.39$ & $0.28 \pm 0.79$ \\
 VJTNN & -  & $3.55  \pm 1.67$ & $2.33  \pm 1.24$ \\
 GCPN & $4.12  \pm 1.19$& $2.49 \pm 1.30$ & $0.79  \pm 0.63$ \\
 GraphAF & $5.00 \pm 1.37$& $3.72 \pm 1.19$ & $1.94 \pm 1.00$ \\
 MIMOSA & $3.48 \pm 1.41$& $3.44 \pm 1.41$ & $\mathbf{3.24 \pm 1.48}$ \\
 \midrule
 LRS & $4.68 \pm 1.48$ & $3.63 \pm 1.23$ & $2.29 \pm 1.12$ \\
 LRS-SC & $5.49 \pm 1.37$ & $4.14 \pm 1.08$& $2.81 \pm 0.92$ \\
 LRS-RBFS & $\mathbf{6.23  \pm 1.28}$  & $\mathbf{4.44  \pm 1.15}$ & $2.80 \pm 1.09$ \\
 \bottomrule
\end{tabularx}
\end{table}

We then evaluate our model in the context of molecule improvement. Given an initial molecule $m_0$ and a threshold $\delta$, the objective is to produce a molecule $m^*$ with maximal score so that the similarity $sim(m^*, m_0)$ between $m^*$ and $m_0$ is larger than $\delta$. As in our baselines, the similarity is defined as the Tanimoto similarity between the Morgan fingerprints with radius of 2 of $m^*$ and $m_0$. We simply set the score of a molecule with $sim(m^*, m_0) < \delta$ to be $-\infty$. 

As a first task, we adopt the same setup as most previous works. More precisely, we optimize the NPlogP scores of 800 molecules with initially low score in ZINC test set, and report the average score improvement for different values of $\delta$. 
The results are presented in table \ref{results:molecule improvement NPlogP}.
Our model outperforms all baselines by a large margin, except MIMOSA in the case $\delta=0.6$.


\begin{table}[t]
\caption{QED and DRD2 score improvement and success for a similarity threshold $\delta = 0.4$.}
\label{results:molecule improvement QED and DRD2}
\begin{tabularx}{\linewidth}{@{}lXXXX@{}}
 \toprule
 & \multicolumn{2}{c}{QED}&\multicolumn{2}{c}{DRD2}\\
 \cmidrule(l){2-3}
 \cmidrule(l){4-5}
 & Improvement & Success& Improvement & Success\\
 \cmidrule(l){1-3}
 \cmidrule(l){4-5}
VJTNN + GAN & $0.11(0.08)$ &  $79.3\%$& $0.61(0.39)$&$\mathbf{97.3\%}$\\
MIMOSA & $0.10(0.05)$&$16.3\%$& $0.28(0.30)$&$25.8\%$\\
 \cmidrule(l){1-3}
 \cmidrule(l){4-5}
LRS-SC& $0.17(0.04)$& $80.0\% $&$0.73(0.31)$&$79.1\%$\\
LRS-RBFS& $\mathbf{0.18(0.03)}$ & $\mathbf{94.5\%}$ &$\mathbf{0.76(0.26)}$&$83\%$\\
 \bottomrule
\end{tabularx}
\end{table}

Then, following \citet{jin2018learning}, we optimize the QED score and the DRD2 score, which is the affinity with dopamine receptor of type 2, as predicted by a model from \citet{olivecrona2017molecular}, and report the average score improvement and success rate. In the former case, we optimize a set of 800 molecules with QED scores between 0.7 and 0.8. An optimization is considered a success if a candidate with a score greater than 0.9 is found. In the latter case, we optimize a set of 1000 molecules with DRD2 scores lower than 0.05. An optimization is considered successful if a candidate with  a score above 0.5 is found. For those experiments, we use a threshold $\delta = 0.4$. The results are presented in table \ref{results:molecule improvement QED and DRD2}.  In general, our method produces larger improvements.

\section{Conclusion}

In this work, we proposed a method for molecular property optimization that only produces realistic candidates with respect to a given dataset. In order to do so, we start by learning the dataset distribution using a generative graph autoregressive model. Based on this learned model, we then perform sampling akin to Metropolis Hastings, implicitly forcing the exploration to focus on regions with higher probability under the training distribution.

We demonstrate that the generated molecules are visually varied and realistic, compared to all reinforcement learning based baselines. Furthermore, our model outperforms all baselines in the task of molecule \emph{improvement}, showing that enforcing realism is not only desirable but also beneficial. 

Notably, the optimization step in our model is not based on learning, so that no data specific to the optimized score is required, which is essential for practical applications where data is scarce.

These results prove that the combination of distribution learning and traditional importance sampling is promising to improve current molecule design methods. Other than Markov Chain Monte Carlo methods, like ours, we believe that similar ideas could be promising when applied to genetic algorithms, which are prominent in modern pipelines. We leave this perspective to future work.

\bibliography{biblio}

\begin{thebibliography}{22}
\providecommand{\natexlab}[1]{#1}
\providecommand{\url}[1]{\texttt{#1}}
\expandafter\ifx\csname urlstyle\endcsname\relax
  \providecommand{\doi}[1]{doi: #1}\else
  \providecommand{\doi}{doi: \begingroup \urlstyle{rm}\Url}\fi

\bibitem[Alperstein et~al.(2019)Alperstein, Cherkasov, and
  Rolfe]{alperstein2019all}
Zaccary Alperstein, Artem Cherkasov, and Jason~Tyler Rolfe.
\newblock All smiles variational autoencoder.
\newblock \emph{arXiv preprint arXiv:1905.13343}, 2019.

\bibitem[Brown et~al.(2019)Brown, Fiscato, Segler, and
  Vaucher]{brown2019guacamol}
Nathan Brown, Marco Fiscato, Marwin~HS Segler, and Alain~C Vaucher.
\newblock Guacamol: benchmarking models for de novo molecular design.
\newblock \emph{Journal of chemical information and modeling}, 59\penalty0
  (3):\penalty0 1096--1108, 2019.

\bibitem[Fu et~al.(2020)Fu, Xiao, Li, Glass, and Sun]{fu2020mimosa}
Tianfan Fu, Cao Xiao, Xinhao Li, Lucas~M Glass, and Jimeng Sun.
\newblock Mimosa: Multi-constraint molecule sampling for molecule optimization.
\newblock \emph{arXiv preprint arXiv:2010.02318}, 2020.

\bibitem[Gao and Coley(2020)]{gao2020synthesizability}
Wenhao Gao and Connor~W Coley.
\newblock The synthesizability of molecules proposed by generative models.
\newblock \emph{Journal of Chemical Information and Modeling}, 2020.

\bibitem[Gilmer et~al.(2017)Gilmer, Schoenholz, Riley, Vinyals, and
  Dahl]{gilmer2017neural}
Justin Gilmer, Samuel~S Schoenholz, Patrick~F Riley, Oriol Vinyals, and
  George~E Dahl.
\newblock Neural message passing for quantum chemistry.
\newblock \emph{arXiv preprint arXiv:1704.01212}, 2017.

\bibitem[Hu* et~al.(2020)Hu*, Liu*, Gomes, Zitnik, Liang, Pande, and
  Leskovec]{Hu2020}
Weihua Hu*, Bowen Liu*, Joseph Gomes, Marinka Zitnik, Percy Liang, Vijay Pande,
  and Jure Leskovec.
\newblock Strategies for pre-training graph neural networks.
\newblock In \emph{International Conference on Learning Representations}, 2020.
\newblock URL \url{https://openreview.net/forum?id=HJlWWJSFDH}.

\bibitem[Irwin and Shoichet(2005)]{irwin2005zinc}
John~J Irwin and Brian~K Shoichet.
\newblock Zinc- a free database of commercially available compounds for virtual
  screening.
\newblock \emph{Journal of chemical information and modeling}, 45\penalty0
  (1):\penalty0 177--182, 2005.

\bibitem[Jin et~al.(2018{\natexlab{a}})Jin, Barzilay, and
  Jaakkola]{jin2018junction}
Wengong Jin, Regina Barzilay, and Tommi Jaakkola.
\newblock Junction tree variational autoencoder for molecular graph generation.
\newblock \emph{arXiv preprint arXiv:1802.04364}, 2018{\natexlab{a}}.

\bibitem[Jin et~al.(2018{\natexlab{b}})Jin, Yang, Barzilay, and
  Jaakkola]{jin2018learning}
Wengong Jin, Kevin Yang, Regina Barzilay, and Tommi Jaakkola.
\newblock Learning multimodal graph-to-graph translation for molecular
  optimization.
\newblock \emph{arXiv preprint arXiv:1812.01070}, 2018{\natexlab{b}}.

\bibitem[Kingma and Welling(2014)]{Kingma2013VAE}
Diederik~P. Kingma and Max Welling.
\newblock Auto-encoding variational bayes.
\newblock In Yoshua Bengio and Yann LeCun, editors, \emph{2nd International
  Conference on Learning Representations, {ICLR} 2014, Banff, AB, Canada, April
  14-16, 2014, Conference Track Proceedings}, 2014.

\bibitem[Krenn et~al.(2020)Krenn, H{\"a}se, Nigam, Friederich, and
  Aspuru-Guzik]{krenn2020self}
Mario Krenn, Florian H{\"a}se, AkshatKumar Nigam, Pascal Friederich, and Alan
  Aspuru-Guzik.
\newblock Self-referencing embedded strings (selfies): A 100\% robust molecular
  string representation.
\newblock \emph{Machine Learning: Science and Technology}, 1\penalty0
  (4):\penalty0 045024, 2020.

\bibitem[Li et~al.(2018)Li, Zhang, and Liu]{li2018multi-objective}
Yibo Li, Liangren Zhang, and Zhenming Liu.
\newblock Multi-{Objective} {De} {Novo} {Drug} {Design} with {Conditional}
  {Graph} {Generative} {Model}.
\newblock \emph{arXiv:1801.07299 [cs, q-bio]}, April 2018.
\newblock URL \url{http://arxiv.org/abs/1801.07299}.
\newblock arXiv: 1801.07299.

\bibitem[Olivecrona et~al.(2017)Olivecrona, Blaschke, Engkvist, and
  Chen]{olivecrona2017molecular}
Marcus Olivecrona, Thomas Blaschke, Ola Engkvist, and Hongming Chen.
\newblock Molecular de-novo design through deep reinforcement learning.
\newblock \emph{Journal of cheminformatics}, 9\penalty0 (1):\penalty0 48, 2017.

\bibitem[Popova et~al.(2019)Popova, Shvets, Oliva, and
  Isayev]{popova2019molecularrnn}
Mariya Popova, Mykhailo Shvets, Junier Oliva, and Olexandr Isayev.
\newblock Molecularrnn: Generating realistic molecular graphs with optimized
  properties.
\newblock \emph{arXiv preprint arXiv:1905.13372}, 2019.

\bibitem[Preuer et~al.(2018)Preuer, Renz, Unterthiner, Hochreiter, and
  Klambauer]{preuer2018frechet}
Kristina Preuer, Philipp Renz, Thomas Unterthiner, Sepp Hochreiter, and
  G{\"u}nter Klambauer.
\newblock Fr{\'e}chet chemnet distance: a metric for generative models for
  molecules in drug discovery.
\newblock \emph{Journal of chemical information and modeling}, 58\penalty0
  (9):\penalty0 1736--1741, 2018.

\bibitem[Segler et~al.(2018)Segler, Kogej, Tyrchan, and
  Waller]{segler2018generating}
Marwin~HS Segler, Thierry Kogej, Christian Tyrchan, and Mark~P Waller.
\newblock Generating focused molecule libraries for drug discovery with
  recurrent neural networks.
\newblock \emph{ACS central science}, 4\penalty0 (1):\penalty0 120--131, 2018.

\bibitem[Shi et~al.(2020)Shi, Xu, Zhu, Zhang, Zhang, and Tang]{shi2020graphaf}
Chence Shi, Minkai Xu, Zhaocheng Zhu, Weinan Zhang, Ming Zhang, and Jian Tang.
\newblock Graphaf: a flow-based autoregressive model for molecular graph
  generation.
\newblock \emph{arXiv preprint arXiv:2001.09382}, 2020.

\bibitem[Simonovsky and Komodakis(2018)]{simonovsky2018graphvae}
Martin Simonovsky and Nikos Komodakis.
\newblock Graphvae: Towards generation of small graphs using variational
  autoencoders.
\newblock In \emph{International Conference on Artificial Neural Networks},
  pages 412--422. Springer, 2018.

\bibitem[Wu et~al.(2018)Wu, Ramsundar, Feinberg, Gomes, Geniesse, Pappu,
  Leswing, and Pande]{wu2018moleculenet}
Zhenqin Wu, Bharath Ramsundar, Evan~N Feinberg, Joseph Gomes, Caleb Geniesse,
  Aneesh~S Pappu, Karl Leswing, and Vijay Pande.
\newblock Moleculenet: a benchmark for molecular machine learning.
\newblock \emph{Chemical science}, 9\penalty0 (2):\penalty0 513--530, 2018.

\bibitem[Xie et~al.(2021)Xie, Shi, Zhou, Yang, Zhang, Yu, and Li]{xie2021mars}
Yutong Xie, Chence Shi, Hao Zhou, Yuwei Yang, Weinan Zhang, Yong Yu, and Lei
  Li.
\newblock Mars: Markov molecular sampling for multi-objective drug discovery.
\newblock \emph{arXiv preprint arXiv:2103.10432}, 2021.

\bibitem[You et~al.(2018)You, Liu, Ying, Pande, and Leskovec]{you2018graph}
Jiaxuan You, Bowen Liu, Zhitao Ying, Vijay Pande, and Jure Leskovec.
\newblock Graph convolutional policy network for goal-directed molecular graph
  generation.
\newblock In \emph{Advances in neural information processing systems}, pages
  6410--6421, 2018.

\bibitem[Zhou et~al.(2019)Zhou, Kearnes, Li, Zare, and
  Riley]{zhou2019optimization}
Zhenpeng Zhou, Steven Kearnes, Li~Li, Richard~N Zare, and Patrick Riley.
\newblock Optimization of molecules via deep reinforcement learning.
\newblock \emph{Scientific reports}, 9\penalty0 (1):\penalty0 1--10, 2019.

\end{thebibliography}
\bibliographystyle{plainnat}

\newpage
\renewcommand{\appendixpagename}{Appendix} \begin{appendices}

\section{Demonstration of Lemma \ref{lemma}}

The transition probability $\mathcal{T}(G_1|G_0)$ defined in section \ref{section:mcmc} is defined by:
\begin{enumerate}
    \item Sampling a sequence $s$ of length $M$ uniformely in $seqs(G_0)$.
    \item Keeping only the $M-\lambda$ first terms of the sequence to define $s^*=s_{M-\lambda}$
    \item Resampling using the autoregressive distribution $\mathcal{P}$ using $s^*$ as initial condition.
\end{enumerate}

In the following, we note $s^*\mathcal{S}$ the ensemble of sequences in $\mathcal{S}$ starting with $s^*$,

The first two steps define a distribution $\mathcal{D}(s^*|G_0)$, which is the probability to sample $s^*$ from the initial graph $G_0$:
\begin{equation}
\begin{aligned}
\label{eq:d(s*|g0)}
    \mathcal{D}(s^*|G_0) &= \sum_{s \in seqs(G_0) \cap s^*\mathcal{S}} \frac{1}{|seqs(G_0)|} \\
    &= \frac{|seqs(G_0) \cap s^*\mathcal{S}|}{|seqs(G_0)|}
\end{aligned}
\end{equation}

Then we define the probability corresponding to the third step. The joint probability over graph and sequence is $\mathcal{P}(G, s) = \mathcal{P}(s)\delta(s \in seqs(G))$ where $\delta(.\in A)$ is the indicator function of the ensemble A. Thus, marginalizing over s:

\begin{equation}
\label{eq:p(g)}
    \mathcal{P}(G) = \sum_{s \in seqs(G)} \mathcal{P}(s)
\end{equation}

The conditional version of the above is :
\begin{equation}
    \mathcal{P}(G, s|s\in s^*\mathcal{S}) = \mathcal{P}(s|s\in s^*\mathcal{S})\delta(s \in seqs(G))
\end{equation}
with
\begin{equation}
   \mathcal{P}(s|s\in s^*\mathcal{S}) = \delta( s \in s^*\mathcal{S})\frac{\mathcal{P}(s)}{\mathcal{P}(s^*)}
\end{equation}
The probability of sampling $G_1$ using $s^*$ as an initial condition is the marginalisation of that expression so that we can define:

\begin{equation}
\label{eq:p(g|s*)}
    \mathcal{P}(G|s\in s^*\mathcal{S}) = \sum_{s \in seqs(G) \cap s^*\mathcal{S}} \frac{\mathcal{P}(s)}{\mathcal{P}(s^*)}
\end{equation}
Abusing notations, we will simply note $\mathcal{P}(G|s^*)$ in the following.

The condition in lemma \ref{lemma} is that $\mathcal{P}(s|G)$ is uniform over $seqs(G)$, so that we can combine equation \ref{eq:p(g)} and \ref{eq:p(g|s*)}:

\begin{equation}
\label{eq:p(g|s*)2}
    \mathcal{P}(G|s^*) = \frac{\mathcal{P}(G)}{\mathcal{P}(s^*)}\frac{|seqs(G) \cap s^*\mathcal{S}|}{|seqs(G)|}
\end{equation}

Finally combining equations \ref{eq:d(s*|g0)} and \ref{eq:p(g|s*)2} we have:

\begin{equation}
\begin{aligned}
    &\mathcal{T}(G_1|G_0) = \sum_{s^*} \mathcal{D}(s^*|G_0)\mathcal{P}(G_1|s^*)\\
    &=\frac{\mathcal{P}(G_1)}{\mathcal{P}(s^*)}\frac{|seqs(G_0) \cap s^*\mathcal{S}| \times |seqs(G_1) \cap s^*\mathcal{S}|}{|seqs(G_0)|\times|seqs(G_1)|}
    \label{eq:t(g1|g0)}
\end{aligned}
\end{equation}

Notice that apart from the $\mathcal{P}(G_1)$, equation \ref{eq:t(g1|g0)} is symmetric upon permutation of $G_1$ and $G_0$. Thus we have:

\begin{equation}
    \mathcal{P}(G_0)\mathcal{T}(G_1|G_0) = \mathcal{P}(G_1)\mathcal{T}(G_0|G_1)
\end{equation}

We can then rewrite the acceptance ratio $r(G_1, G_0)$ defined in equation~\ref{eq:acceptance} as:
\begin{equation}
\begin{aligned}
    r(G_1, G_0) &= \frac{F(G_1)}{F(G_0)}\frac{\mathcal{P}(G_1)\mathcal{T}(G_0|G_1)}{{T}(G_1|G_0)} \\ &=\frac{\pi(G_1)\mathcal{T}(G_0|G_1)}{\pi(G_0)\mathcal{T}(G_1|G_0)}
    \label{eq:real_ratio}
\end{aligned}
\end{equation}

where $\pi(G)=F(G)\mathcal{P}(G)$ and equation~\ref{eq:real_ratio} defines a correct acceptance ratio to perform Metropolis Hasting sampling of a $d\propto\pi$, using the asymmetric proposal distribution $\mathcal{T}$.

\section{Implementation details}

\subsection{Input format}
Molecules are presented in kekulized form with implicit hydrogens as graphs with atoms as nodes and chemical bonds as edges. Thus, edges are labelled as SINGLE, DOUBLE, TRIPLE or NO EDGE. 

Concerning the atomic information, in order to fully describe the molecule (except for the chirality), we retain the information about the atomic number, its valence (counting hydrogen) and charge. We find that there are 20 different cases occuring in the ZINC dataset, to which we add the special STOP NODE case. Thus, based on that information, the initial node embedding is a one hot embedding of dimension 21.

\subsection{Training details and hyperparameters}

\subsection{Hardware}
All experiments have been run on a Geforce GTX 1080 Ti. The autoregressive models have been trained until convergence, which takes 2 to 3 days. We manually tuned the hyperparameters of each model. No fine tuning algorithm has been used, and the final hyperparameters are given in supplementary materials. In the distribution learning section, we report the score of the model of each type that we use in the optimization sections.

\subsubsection{Model hyperparameters and training}
Our model consists in three layers of convolution as described in equation \ref{eq:conv} followed by two MLP with $Tanh$ non linearity except on the last layer are used as $MLP^{node}$ and $MLP^{edge}$ in equation \ref{eq:mlp}. Those MLPs have 8 layers in the case of LRS and LRS-SC, and 5 layers in the case of LRS-RBFS. 
The network is trained by minimizing cross entropy loss, using the Adam optimizer with a learning rate of $1e-3$ , using the \emph{Pytorch} and \emph{Pytorch-geometric} librairies.

\subsubsection{Optimization hyperparameters}
Optimization is performed using an initial temperature $T_0=0.1$ in equation \ref{eq:acceptance} with a cooling rate of 0.1 per 100 iterations. Each optimization is taken has the best score among $R$ runs of $S$ steps of the algorithm. For NPLogP and QED, we set $R=10$, while for DRD2 we set $R=20$. We set $S=50$ for NPLogP and $S=150$ for QED and DRD2.

\section{Training a network conditioned on size}
Training a network conditioned on size as defined in section \ref{section:size} will generally lead to a disappointing local optimum with relatively high loss $\widehat{L}$, where the network perfectly matches the expected size but produces qualitatively poor molecules. 
We found that we could perform training for a few epochs with $C=0$, until the loss is lower than $\widehat{L}$, and then continue the training allowing C to vary.

\section{Metrics}
\label{metrics definition}
All metrics are evaluated on a sample $\mathcal{G}$ of 10,000 generated molecules. The similarity $sim(m, m')$ between the molecules $m$ and $m'$ is defined as Tanimoto similarity between ECFP6 fingerprints of the two molecules.

\subsection*{Validity}
A molecule is considered valid if it can be parsed as a valid SMILES with the RDKIT librairy. Non connexe molecules are \emph{not} considered valid. The validity score is the ratio of valid molecules among 10,000.

\subsection*{Novelty}
The novelty score is the ratio of \emph{valid} molecules that do not appear in the dataset over all \emph{valid} generated molecules.

\subsection*{Uniqueness}
The uniqueness score is the ratio of different \emph{valid} generated molecules over all \emph{valid} generated molecules.

\subsection*{Fréchet ChemNet Distance}
We use the opensource fcd\_torch implementation of the Fréchet ChemNet Distance.

\subsection*{KL score}
We use an similar definition of the KL score as \cite{brown2019guacamol}. More precisely, the dataset and generated molecule distributions, $d_1^p$ and $d_2^p$ are evaluated on a set of property descriptors $p$ and the standard Kullback-Leibler divergence is measured between the two distributions. The average is then taken on $p$. The only difference with \cite{brown2019guacamol} is that we directly provide this quantity while they apply an exponential to map the score between 0 and 1.
The descriptors are properties from rdkit:
\begin{itemize}
    \item The BertzCT index
    \item the molecule LogP
    \item The molecular weight
    \item The total polar surface area
    \item The number of donors
    \item The number of acceptors
    \item The number of rotatable bonds
    \item The number of aliphatic rings
    \item The number of aromatic rings
    \item The internal pairwise similarity
\end{itemize}
The distribution of continuous properties is evaluated using a gaussian kernel density estimator.

\subsection{Log-Likelihood}
In section \ref{section: unconstrained optimisation}, we provide the log likelihood of top 3 generated molecules in feature space. 

In ChemNet feature space, we evaluate the dataset distribution with a multivariate gaussian, in a similar fashion than the FCD in calculated.  

In descriptor features space, we use the same list of descriptor as for the KL metric, except for the molecule LogP (which is the optimized quantity). For the discrete properties, a basic frequency estimator fails to be informative, as the log likelihood becomes $-\infty$ for molecules which have features never seen in the dataset, without making a difference between small and large deviations. For example, GraphAF produces molecules with up to 40 rotatable bonds, while MRNN or LRS-RBFS produce 11 and 13 rotatable bonds. There is never more than 10 rotatable bonds in the dataset, so that all three models produce $-\infty$ lo likelihood.

To circumvent the issue, we simply estimate \emph{all} properties with a Gaussian distribution, and report the likelihood accordingly.

\section{Baseline Models}

The code from JTVAE, GCPN and GraphAF is available and was used to produce the results in the distribution learning section. The code from MRNN is not officially available and we report the result from the original paper. The SMILES LSTM model implementation is the hill climbing version from the Guacamol baselines by \cite{brown2019guacamol}.

In the unconstrained optimization section, we reported the results for the molecules given in the original papers. The only case where it's not possible is the SMILES LSTM model, which we rerun for this specific task.

In the molecule improvement section, we report the results from the original papers, with a few exceptions. The results for JTVAE are the updated one reported in their following paper\citep{jin2018learning}. All models use the same list of SMILES, that can be obtained from he VJTNN implementation by \cite{jin2018learning}. The original results of GraphAF did not use the same list of SMILES to improve, we reran their original code with the correct list of SMILES. Finally, the MIMOSA model did not penalize the similarity as done in this work. In their original implementation, the similarity is added to the score, ensuring that mostly similar candidate are generated. We modified their implementation to remove that feature and reran their code with the same hyperparameters.
\end{appendices}
\end{document}